# Online Physical Enhanced Residual Learning for Connected Autonomous Vehicles Platoon Centralized Control

Hang Zhou, Heye Huang\*, Peng Zhang, Haotian Shi, Keke Long, Xiaopeng Li\*

*Abstract—* **This paper introduces an online physical enhanced residual learning (PERL) framework for Connected Autonomous Vehicles (CAVs) platoon, aimed at addressing the challenges posed by the dynamic and unpredictable nature of traffic environments. The proposed framework synergistically combines a physical model, represented by Model Predictive Control (MPC), with data-driven online Q-learning. The MPC controller, enhanced for centralized CAV platoons, employs vehicle velocity as a control input and focuses on multi-objective cooperative optimization. The learning-based residual controller enriches the MPC with prior knowledge and corrects residuals caused by traffic disturbances. The PERL framework not only retains the interpretability and transparency of physics-based models but also significantly improves computational efficiency and control accuracy in real-world scenarios. The experimental results present that the online Q-learning PERL controller, in comparison to the MPC controller and PERL controller with a neural network, exhibits significantly reduced position and velocity errors. Specifically, the PERL's cumulative absolute position and velocity errors are, on average, 86.73% and 55.28% lower than the MPC's, and 12.82% and 18.83% lower than the neural network-based PERL's, in four tests with different reference trajectories and errors. The results demonstrate our advanced framework's superior accuracy and quick convergence capabilities, proving its effectiveness in maintaining platoon stability under diverse conditions.**

## I. INTRODUCTION

Connected autonomous vehicles (CAVs) platoon with inter-vehicle communication permits vehicles to travel close together, enhancing road capacity and traffic safety [1], [2]. The vehicle platoon system integrates a complex network of interconnected agents, involving multiple vehicles. The system is characterized by the non-linearity and coupling of individual vehicle dynamics models, the interactions between system agents, model uncertainties, and external disturbances, all of which can pose significant challenges to the performance of platoon control [3], [4], as illustrated in **Figure 1**. Considering real-world scenarios where vehicle platoons interact with dynamic traffic under unpredictable conditions, including extreme weather like rain or snow, maintaining steady-state control is crucial for safe destination arrival. Therefore, a safe and precise motion controller is essential for achieving the stability of CAVs formation, enhancing operability, and ensuring robustness against interferences.

Research supported by the U.S. National Science Foundation under Grant No.2313578. (\*corresponding author: Heye Huang, Xiaopeng Li).

Hang Zhou, Heye Huang, Peng Zhang, Haotian Shi, Keke Long, and Xiaopeng Li, are now with the Department of Civil and Environmental Engineering, Univ. of Wisconsin-Madison, Madison, WI, USA. (e-mails: hhuang468@wisc.edu; xli2485@wisc.edu).

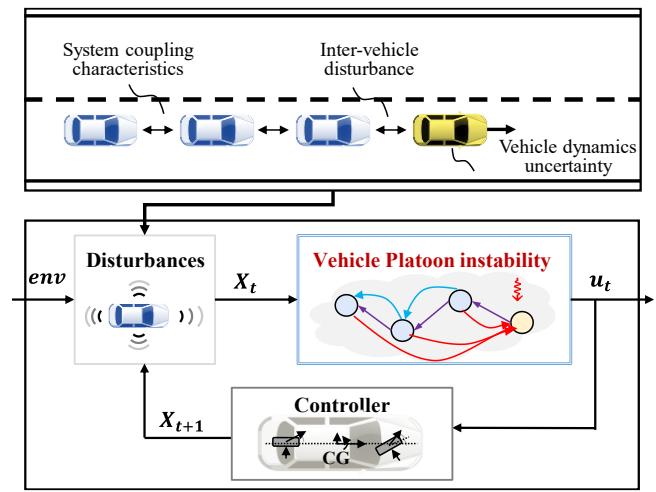

Figure 1. Safety control for vehicle platoon under dynamic disturbances.

Existing methods for CAVs platoon control can be categorized into two categories [5], [6]: model-based methods (e.g. rule-driven, optimization methods), and learning-based methods (e.g. multi-agent collaborative control based on reinforcement learning and deep learning). Many platoon control designs have focused on classic control approaches based on physics models [7]. Leveraging a robust theoretical foundation, these methods provide a solid understanding and control of vehicle dynamics, offering universally applicable modeling, control, and analytical solutions for autonomous vehicle platoon control. Specifically, static linear controllers represent one of the most thoroughly investigated methods [8], [9]. They are convenient for application and facilitate the establishment of closed-loop system models for theoretical analysis of various system performances. However, this type of controller struggles to support constrained optimization frameworks with multiple explicit objectives and constraints. Furthermore, methods like Sliding Mode Control (SMC) [10], adaptive control, and Model Predictive Control (MPC) [11] have also been developed. Particularly, MPC has garnered attention due to its ability to explicitly handle safety constraints, integrate optimization of collaborative control objectives, and its clear potential for distributed application and robustness [4], [12]. These traditional methods often linearize complex systems to facilitate theoretical study, which tends to overly simplify complex dynamics and overlook the dynamic variability of platoon systems. This leads to model-based networked autonomous vehicle control methods facing challenges in effective application within real traffic.

Recently, data-driven machine learning methods [13], [14], including Deep Learning (DL) and Reinforcement Learning (RL), have emerged as formidable tools [15], [16]. DL-based or data-driven model-free control methods, independent of dynamical models, can leverage driving trajectory data to directly design control strategies for networked autonomous vehicles. These methods are capable of capturing complex nonlinear relationships within data, enabling effective handling of various driving scenarios. RL is a commonly used model-free control method in the control of intelligent networked vehicles in mixed traffic. Various training algorithms, such as Deep Q-Networks (DQN) [17], [18] and Deep Deterministic Policy Gradients (DDPG) [19], [20], have been widely applied. Sallab et al. [21] employed deep reinforcement learning to implement lane-keeping control in the open racing car simulator (TORCS), comparing the discrete-space DQN method with the continuous action space DDAC method, demonstrating the latter's ability to achieve excellent control effects and smooth trajectories. Shi et al. [16] introduced a DRL-based cooperative CAV longitudinal control strategy for mixed traffic settings, segmenting the mixed platoon into multiple subsystems for efficient centralized cooperative control. Furthermore, multi-agent reinforcement learning, a concept explored by Busoniu et al. [22], has been widely adopted in networked CAVs platoon control [23], [24]. Compared to other machine learning algorithms, the reinforcement Q-learning method in cooperative control scenarios enables direct and simple output of Q-values in the current state to choose the best action sequence [25]. With sufficient samples or observational data, Q-learning learns the optimal state-action pairs. In practice, it has been proven to converge to the optimal state-action value. The Q-network reinforcement learning technique in [26] determines the optimal locations for base stations to provide enhanced platoon features to CAVs. Given the high dimensionality, continuous state and action spaces, and non-linearity of networked autonomous vehicle control problems, RL-based control strategies can learn complex control models through continuous exploration of the environment [27], [28]. However, they often lack interpretability and transparency, making understanding the control processes and dynamic mechanisms of multi-vehicle autonomous driving challenging [29]. Additionally, the data collection process for the required training data is inherently risky, with models only being usable post-pretraining. To summarize, both physical models and learning methods alone are inadequate for the complex CAVs platoon control problem, which will cause the failure of safe and precise control, especially under special conditions.

To address this gap, this paper introduces an online learning physical enhanced residual learning (PERL) controller for centralized CAV platoons. This framework integrates the physical model with data-driven RL techniques. The vehicle dynamics' physical model, represented by the centralized MPC, provides prior knowledge and safe constraints, while Q-learning, employed as an online residual learning method, focuses on capturing additional errors, such as those stemming from incorrect calibration of the physical model, thereby refining the model's output. By integration, the controller bridges the gap between theoretical foundations and the dynamic, complex realities of autonomous driving scenarios. Our contributions are as follows:

- We develop a novel PERL framework for CAVs platoon control, maintaining the inherent interpretability and transparency of physics-based models. This framework simplifies troubleshooting and fine-tuning and theoretically demonstrates the control accuracy and stability.
- An enhanced centralized MPC controller is formalized for CAVs platoon. The velocity is employed as the control input for CAVs and considers multi-objective cooperative optimization, improving actual computational efficiency.
- We integrate a physical model with a data-driven online residual learning model. The physical MPC imbues PERL with a priori knowledge and the Q-learning composites residuals of the physical model. Experimental results indicate its commendable accuracy and rapid convergence.

The rest of this paper is organized as follows. Section 2 introduces our proposed PERL control method from the aspect of the framework, physical MPC controller, and the residual learning component Q-learning. In Section 3, we describe the experiment condition setting and present quantitative experiment results. Conclusions are given in Section 4.

II. METHODOLOGY

*A. Model Framework*

The framework of the proposed PERL controller is illustrated in **Figure 2**, consisting of two modules: 1) Fundamental physical-based control. We employ the MPC-based controller, which considers platoon constraints for centralized control. The objective of our controller is to achieve the desired platoon acceleration, velocity, and individual inter-vehicle distances while ensuring safety constraints. 2) Learning-based residual control. A residual feedback module is integrated into the traditional physical model. A reinforcement learning method Q-learning is utilized for online learning, fitting, and compensation of system model errors and external disturbances. This allows for appropriate driving speeds and real-time control output adjustments, enabling vehicles to minimize deviation from the target trajectory and maximize stability.

To help readers understand the relationship between these two modules, we introduce the workflow of the platoon control under the PERL controller before introducing the two modules. Consider an environment with discrete time steps. As illustrated in **Figure 2**, at the beginning of time step $k$, the platoon detects the information of the current state $X_k$. Then, the physical model is applied to obtain the optimal desired control input $u_k^p = f^{\text{MPC}}(X_k)$. After that, the residual learning component runs with the inputs $u_k^p$ to obtain the desired control input with residual $u_k^r = f^{\text{RL}}(u_k^p)$. Finally, the controller applies $u_k^r$ to the platoon. The state $X_k$ transfers to the next stage $X_{k+1}$ based on the real action with the system model errors and external disturbances $u_k^a$. In the next two sections, we will discuss the details of the physical MPC controller and the online Q-learning method.

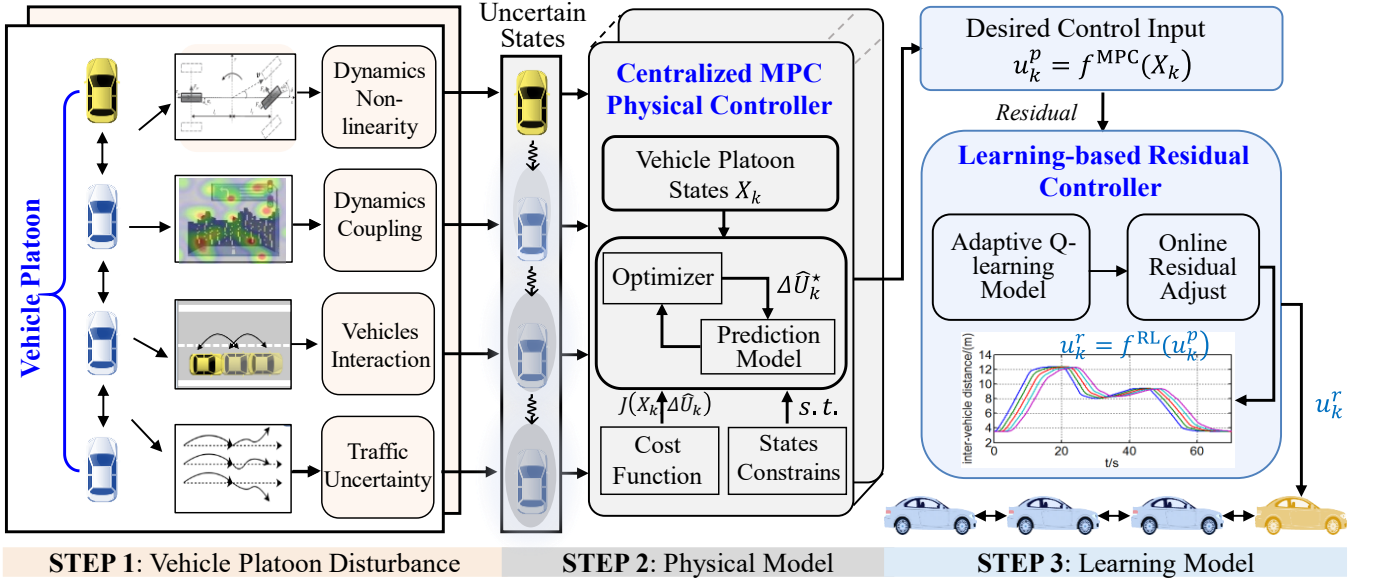

Figure 2. Illustration of the integrated CAVs platoon control framework. The inputs to the controller are uncertain vehicle platoon states with multiple disturbances. The output is the control action with residual compensation for all vehicles in the platoon.

*B. Centralized Vehicle Platoon Control with MPC*

Centralized control necessitates that the central controller solves for the optimal control of each vehicle at every timestep. Addressing this type of centralized control involves two key challenges: (1) the state and control of all vehicles are intertwined through the objective function and constraints; (2) a longer planning horizon requires forecasting future traffic dynamics, which can be impacted by the curse of dimensionality and disturbances.

The platoon as shown in **Figure 3**, comprises $I + 1$ homogeneous vehicles, including a leading vehicle (the leader) and $I$ following vehicles (the followers). Each vehicle in the platoon is an integral part of the control system. Operating under a predefined reference trajectory, the goal of the MPC controller is not just to approximate the real trajectory to this reference, but to do so with a level of precision that maximizes overall system efficiency, which ensures the smooth and safe operation of the entire platoon.

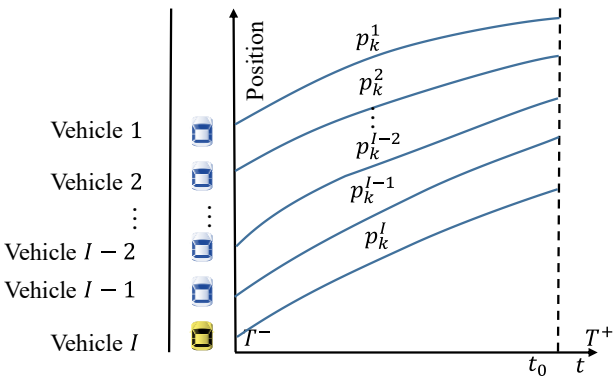

Figure 3. The centralized connected autonomous vehicles platoon.

According to the vehicle longitudinal dynamics, the linear model for a single vehicle $i \in \mathcal{I} = \{1, 2, \ldots, I\}$ at time step $k \in \mathbb{Z}$ with constant sampling interval $\Delta t$:

$$\begin{cases} p_{k+1}^i = p_k^i + v_k^i \Delta t + \frac{1}{2} a_k^i \Delta t^2 \\ v_{k+1}^i = v_k^i + a_k^i \Delta t \\ a_{k+1}^i = -\frac{\Delta t}{\tau^i} v_k^i + \frac{\Delta t}{\tau^i} u_k^i \end{cases} \quad (1)$$

where $p_k^i$ denotes the position, $v_k^i$ denotes the speed, $a_k^i$ denotes the acceleration, $u_k^i$ is the control input or desired speed of vehicle $i$ at time step $k$, and $\tau^i$ is the inertial delay of vehicle longitudinal dynamics.

Based on this, we can obtain the state space model for vehicle $i$ at time $k$:

$$x_{k+1}^i = A^i x_k^i + B^i u_k^i \quad (2)$$

where $x_k^i = [p_k^i, v_k^i, a_k^i]$ is the state information for vehicle $i$, and

$$A^i = \begin{bmatrix} 1 & \Delta t & \frac{\Delta t^2}{\tau^i} \\ 0 & 1 & \Delta t \\ 0 & -\frac{\Delta t}{\tau^i} & 0 \end{bmatrix}, B^i = \begin{bmatrix} 0 \\ 0 \\ \frac{\Delta t}{\tau^i} \end{bmatrix} \quad (3)$$

We assume that all vehicles in the platoon are homogeneous ($A^i = A, B^i = B, \tau^i = \tau$). Then we define the state variables and the control inputs for:

$$X_k = [x_k^1, \ldots, x_k^I]^\top = [p_k^1, \ldots, p_k^I, v_k^1, \ldots, v_k^I, a_k^1, \ldots, a_k^I]^\top \quad (4)$$
$$U_k = [u_k^1, \ldots, u_k^I]^\top \quad (5)$$

such that the platoon dynamics are:

$$X_{k+1} = A_I X_k + B_I U_k \quad (6)$$

where $A_I = A \otimes E_I, B_I = B \otimes E_I, \otimes$ is the Kronecker operator, and $E_I$ is the $I$ dimensional elementary matrix.

Define the change in control actions $\Delta U_k$ from the previous control action $U_{k-1}$:

$$\Delta U_k = U_k - U_{k-1} = [\Delta u_k^1, \ldots, \Delta u_k^I]^\top \quad (7)$$

where $\Delta u_k^i$ is the change in control action for vehicle $i$.

To predict the state value of the platoon for the next $N$ steps, we introduce the predicted state value of the platoon at time $k + n$ for $n \in \mathcal{N} = \{1, \ldots, N\}$ from the measured state value at time $k$ using the model denoted as $\hat{X}_{k+n|k}$, with the prediction window defined as:
$$\mathcal{X}_k = [\hat{X}_{k+1|k}^\top, \ldots, \hat{X}_{k+N|k}^\top]^\top \tag{8}$$
for the predicted value of the platoon controller as:
$$\Delta \hat{U}_k = [\Delta \hat{U}_{k|k}^\top, \ldots, \Delta \hat{U}_{k+N-1|k}^\top]^\top \tag{9}$$
where from the measurement at time $k$ the predicted applied control at time $k + n$ is:
$$\hat{U}_{k+n|k} = \hat{U}_{k+n-1|k} + \Delta \hat{U}_{k+n|k} \tag{10}$$
$$\Delta \hat{U}_{k+n|k} = [\Delta \hat{u}_{k+n|k}^1, \ldots, \Delta \hat{u}_{k+n|k}^I]^\top \tag{11}$$

The state prediction of the platoon $\mathcal{X}_k$ can be written as a linear combination of the current state $X_k$, the previously applied control $U_{k-1}$ and the predicted change in control $\Delta \hat{U}_k$:
$$\mathcal{X}_k = \Phi X_k + \lambda U_{k-1} + \Gamma \Delta \hat{U}_k \tag{12}$$
where
$$\Phi = \begin{bmatrix} A_I \\ \vdots \\ A_I^N \end{bmatrix}, \lambda = \begin{bmatrix} A_I^0 B_I \\ \vdots \\ (A_I^{N-1} + \cdots + A_I^0) B_I \end{bmatrix} \tag{13}$$
$$\Gamma = \begin{bmatrix} B_I & \cdots & 0 \\ \vdots & \ddots & \vdots \\ (A_I^{N-1} + \cdots + A_I^0) B_I & \cdots & B_I \end{bmatrix} \tag{14}$$

Denote the reference state as $p_k^{i\star}, v_k^{i\star}, a_k^{i\star}$, and
$$X_k^\star = [p_k^{1\star}, \ldots, p_k^{I\star}, v_k^{1\star}, \ldots, v_k^{I\star}, a_k^{1\star}, \ldots, a_k^{I\star}]^\top \tag{15}$$
$$\mathcal{X}_k^\star = [(X_{k+1}^\star)^\top, \ldots, (X_{k+N}^\star)^\top]^\top \tag{16}$$

Consider for each vehicle $i \in \mathcal{I}$, the absolute position, velocity, and acceleration errors as the difference between the current state and the reference state:
$$\begin{cases} \tilde{p}_k^i = p_k^i - p_k^{i*} \\ \tilde{v}_k^i = v_k^i - v_k^{i*} \\ \tilde{a}_k^i = a_k^i - a_k^{i*} \end{cases} \tag{17}$$

For the entire platoon, these errors can be written as:
$$X_k - X_k^* = [\tilde{p}_k^1, \ldots, \tilde{p}_k^I, \tilde{v}_k^1, \ldots, \tilde{v}_k^I, \tilde{a}_k^1, \ldots, \tilde{a}_k^I]^\top \tag{18}$$

Then denote $\hat{p}_{k+n|k}^i, \hat{v}_{k+n|k}^i, \hat{a}_{k+n|k}^i$ as the prediction error where the subscript indicates the state prediction at time $k + n$ given the state at time $k$.

The formulation for the MPC controller with a finite prediction horizon of $N$ steps is:
$$J(k, N) = \min \sum_{n=0}^{N-1} \left[ \sum_{i=1}^{I} q_1 (\hat{p}_{k+n|k}^i)^2 + q_2 (\hat{v}_{k+n|k}^i)^2 + q_3 (\hat{a}_{k+n|k}^i)^2 + q_4 (\Delta \hat{u}_{k+n|k}^i)^2 \right] \tag{19}$$
$s.t.$:
$$d_{\min} \le p^{i-1} - p^i \le d_{\max}, \quad \forall i \in \mathcal{I} \tag{20}$$
$$v_{\min} \le v^i \le v_{\max}, \quad \forall i \in \mathcal{I} \tag{21}$$
$$a_{\min} \le a^i \le a_{\max}, \quad \forall i \in \mathcal{I} \tag{22}$$
where $q_1, q_2, q_3, q_4$ are the penalty on absolute position error, velocity error, acceleration error, and the control inputs, respectively. $d_{\max}$ is the maximum distance for the platoon, and $d_{\min}$ is the minimum value of the safety distance. $a_{\max}, a_{\min}, v_{\max}, v_{\min}$ are the maximum and minimum values of the acceleration and the velocity. Constraints (20) represent the safety and maximum distance of the platoon. Constraints (21) represent the road speed limit.

Constraints (22) represent the acceleration limit based on the engine and the braking systems of the vehicles.

The problem can be written in the form of a quadratic program:
$$J(X_k, \Delta \hat{U}_k) = \Delta \hat{U}_k^\top (\Psi + \Gamma^\top \Omega \Gamma) \Delta \hat{U}_k + 2(\Phi X_k + \lambda U_{k-1} - \mathcal{X}_k^\star)^\top \Omega \Gamma \Delta \hat{U}_k \tag{23}$$
$s.t.$:
$$\bar{G} \Gamma \Delta \hat{U}_k \le -\bar{G}(\Phi X_k + \lambda U_{k-1}) - \bar{g} \tag{24}$$
where
$\Omega = \text{diag}\{Q, \ldots, Q, 0\}, \Psi = \text{diag}\{R_\Delta, \ldots, R_\Delta\}$ are block diagonal matrices,
$$Q = \begin{bmatrix} q_1 E_I & 0 & 0 \\ 0 & q_2 E_I & 0 \\ 0 & 0 & q_3 E_I \end{bmatrix}, R_\Delta = q_4 E_I \tag{25}$$
$$\bar{G} = \text{diag}[\check{G}, \ldots, \check{G}], \bar{g}^\top = [g^\top, \ldots, g^\top], \tag{26}$$
$$\check{G} = \begin{bmatrix} \mathfrak{T}_I & 0 & 0 \\ -\mathfrak{T}_I & 0 & 0 \\ 0 & -E_I & 0 \\ 0 & E_I & 0 \\ 0 & 0 & -E_I \\ 0 & 0 & E_I \end{bmatrix}, g = \begin{bmatrix} 1_{I-1} d_{\min} \\ -1_{I-1} d_{\max} \\ 1_I v_{\min} \\ -1_I v_{\max} \\ 1_I a_{\min} \\ -1_I a_{\max} \end{bmatrix}, \tag{27}$$

$\mathfrak{T}_I$ is a size $(I - 1) * I$ Toeplitz matrix with $-1$ on the diagonal and 1 on the first upper diagonal, and $1_{I-1}$ and $1_I$ are column vectors of ones of size $(I - 1)$ and $I$, respectively. State-of-the-art solvers can quickly solve this quadratic optimization problem.

The optimal platoon control action is the change in control that minimizes the constrained finite horizon cost function
$$\Delta \hat{U}_k^\star = \arg\min_{\Delta \hat{U}_k} J(X_k, \Delta \hat{U}_k) \tag{28}$$
where the first element $\Delta \hat{U}_{k|k}^\star$ will be the output control $u_k^p$ of the physical model.

*C. Online Resudial Learning*

Residual learning aims to approximate the residual term or "gap" between the predicted and actual system states and adjust the control output obtained by the physical model.

In residual learning, the key aspect is to measure the difference between the desired input speed set by the MPC and the actual output speed of the vehicles. This difference guides the adjustment of the vehicle's Direct Control Variable (DCV), which varies with the experimental platform. It's important to note that the input for the DCV changes based on the experimental platform used, which means the output of residual learning must be adapted to fit the actual platform. For instance, in a full-sized AV, the DCV is throttle/brake, while in a reduced-scale robot car, it's the motor's RPM. Our residual learning is expected to capture the disturbance caused by different DCVs.

This paper utilizes Q-learning as the residual learning method. In the design of Q-learning, defining states and actions appropriately is crucial. Considering the structure of the residual learning previously outlined. The state and the action in the Q-learning are defined by the control output $u_k^p$ obtained from the MPC and the changing rate $\eta$ of the control output. Since both the state and action in this scenario are continuous variables, the resulting Q-table would be infinitely large, making it impractical to train. To address this challenge, the study employs two approaches for handling these continuous variables.

For actions, given that the vehicle acceleration is restricted within a limited range $[a_{\min}, a_{\max}]$, the output DCV also has a corresponding range. Therefore, the DCV can be discretized directly. The discretization interval $\Delta$ is a pre-determined parameter. Regarding states, the speed control error is transformed within the range of $[-\sigma, +\sigma]$ using a sigmoid function. Then, fuzzy logic is applied to discretize the continuous state space. This is achieved through the definition of several state membership functions, which help to categorize the continuous states into discrete groups.

In the training phase, the Q-learning agent chooses actions based on predicted maximum rewards, defined by the discrepancy between the vehicle's actual and input speeds; smaller errors yield higher rewards. The Q-table is updated continuously through exploration and interaction until convergence, signifying minimal changes in the table values.

III. DATA EXPERIMENT AND RESULTS

In this section, the proposed PERL controller model's performance is evaluated in a simulation environment through a comparative analysis with two baseline models.

A. Experiment Condition Setting

In this simulation scenario, we set up a platoon consisting of 5 vehicles ($I = 4$) over a duration of $T = 15$ seconds, with a time step of $\Delta t = 0.1$ seconds. Two types of reference trajectories are considered. The first scenario involves uniform motion, where all vehicles maintain a constant speed of 15 m/s. The second scenario involves variable speed, divided into four phases: initially, vehicles proceed at a constant velocity of 15 m/s for 2 seconds, followed by a deceleration at -2 m/s² for 5.5 seconds, then accelerate at 2 m/s² for 5.5 seconds, and finally move at a constant velocity for the remaining 2 seconds. In both scenarios, the initial spacing between the five vehicles is set at 20 m. A minimum safety distance $d_{\max} = 15$ m and the maximum spacing $d_{\min} = 30$ m. To simulate the error that happens during the transfer of the RPM and the velocity, we consider two types of error for the actual control output $u_k^a$. The first one is an affine error $u_k^a = 1.1 u_k + 0.1 + x\ (m/s^2), x \sim N(0, 0.3)$, where $u_k$ is the control input. The second one is quadratic error $u_k^a = 0.01 u_k^2 + u_k + 0.1 + x\ (m/s^2), x \sim N(0, 0.3)$. For the online learning procedure, the residual learning is updated each 20 time steps (i.e., 2s) using the collected data during the experiments.

B. Baseline Models

To make a comprehensive comparison with current prediction models, we compare the performance of three different methods tested: 1. using only MPC, 2. using neural network for residual learning, and 3. using Q-learning for residual learning. All three methods will be evaluated in the two scenarios with two types of errors, i.e., in total four tests. A Multi-Layer Perceptron model with ReLU as an activation function is used as the neural network baseline model. The input and output of the network are the same as described above, i.e., the control output obtained from the MPC and the adjusted control output for the vehicles. Before the online learning, the NN model will be initially trained to make the input the same as the output.

C. Experiment Results

The time-space diagrams of the three methods in Scenario 2 are illustrated in **Figure 4**. The local magnification of the trajectory reveals that under the MPC controller, there is a considerable deviation between the actual trajectory and the reference trajectory. In contrast, with the online PERL method, the actual trajectory closely aligns with the reference trajectory, demonstrating the control accuracy. To further compare the performance of the three control methods, particularly to distinguish between the use of Q-learning and a neural network as the residual learning module in online PERL, we conduct an additional comparative analysis in **Table 1** and **Figure 5**. Given that the primary control output is velocity, we recorded both velocity and position errors to evaluate the performance of these methods. Notice that all the units for position are meters, and for velocity are meters per second.

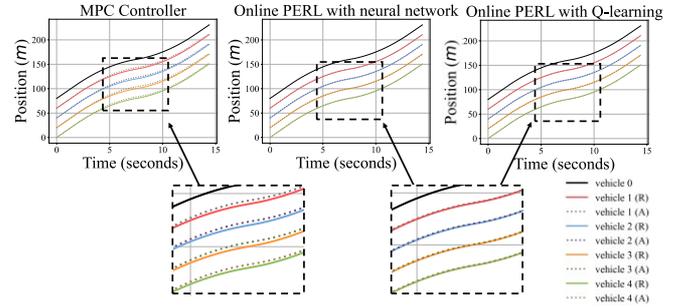

Figure 4. The time-space diagram for the three methods in Scenario 2.

**Table 1** details the cumulative and maximum absolute velocity errors for the three methods across four tests. The results indicate that both variations of the PERL method, which utilize the Q-learning algorithm and a neural network, outperform using the MPC alone in four tests. Specifically, in the tests, the average cumulative absolute velocity errors for the MPC combined with Q-learning are 55.28% smaller compared to using MPC alone. Moreover, the average gap in cumulative absolute position errors between these two methods is 86.73%, showing that the PERL method performance is much better than the MPC controller. When comparing the MPC combined with a neural network, the MPC combined with Q-learning has lower cumulative absolute velocity and position errors in all tests, averaging 12.82% and 18.83%, respectively. From the perspective of maximum absolute error, both the error gaps for position and velocity between the MPC combined with Q-learning and using MPC alone are still large in scenario 2, which are 88.33% and 71.24%, respectively. However, in scenario 1, the error gap of all three methods is smaller than 5%. The similarity arises from limited training data in early online learning stages, where residual learning doesn't adjust control output. In the complex environment of scenario 2, larger error gaps between the other methods and the MPC-Q-learning combination highlight the need for precise control in complex settings.

The variations of the velocity error for the whole trajectories are illustrated in **Figure 5,** which explains the results in **Table 1**. From **Figure 5**, it is evident that in the four tests, all three methods initially exhibit similar trends of the velocity error, ranging between 0.2-0.3. While as training progresses, residual learning learns the error, reducing the velocity error below zero, thereby compensating for the initial position error caused by the positive velocity error at the beginning. Moreover, **Figure 5** also demonstrates that, in all four tests, both PERL controllers eventually stabilize near zero, confirming their robustness and effectiveness. During the uniform motion in scenario 1, it is observed that the velocity error convergence for the MPC combined with Q-learning is noticeably smoother, achieving a minimum error around 0.1, whereas for the MPC combined with a neural network, the minimum error reaches approximately 0.2. Consequently, Q-learning, as an online residual learning approach, demonstrates a faster convergence rate, making it preferable for online PERL methods.

However, the current simulations, using simplified control environments and error forms, may not mirror real-world conditions accurately, thus necessitating further real-world testing to validate the PERL methods' performance, e.g. in the real vehicle platforms.

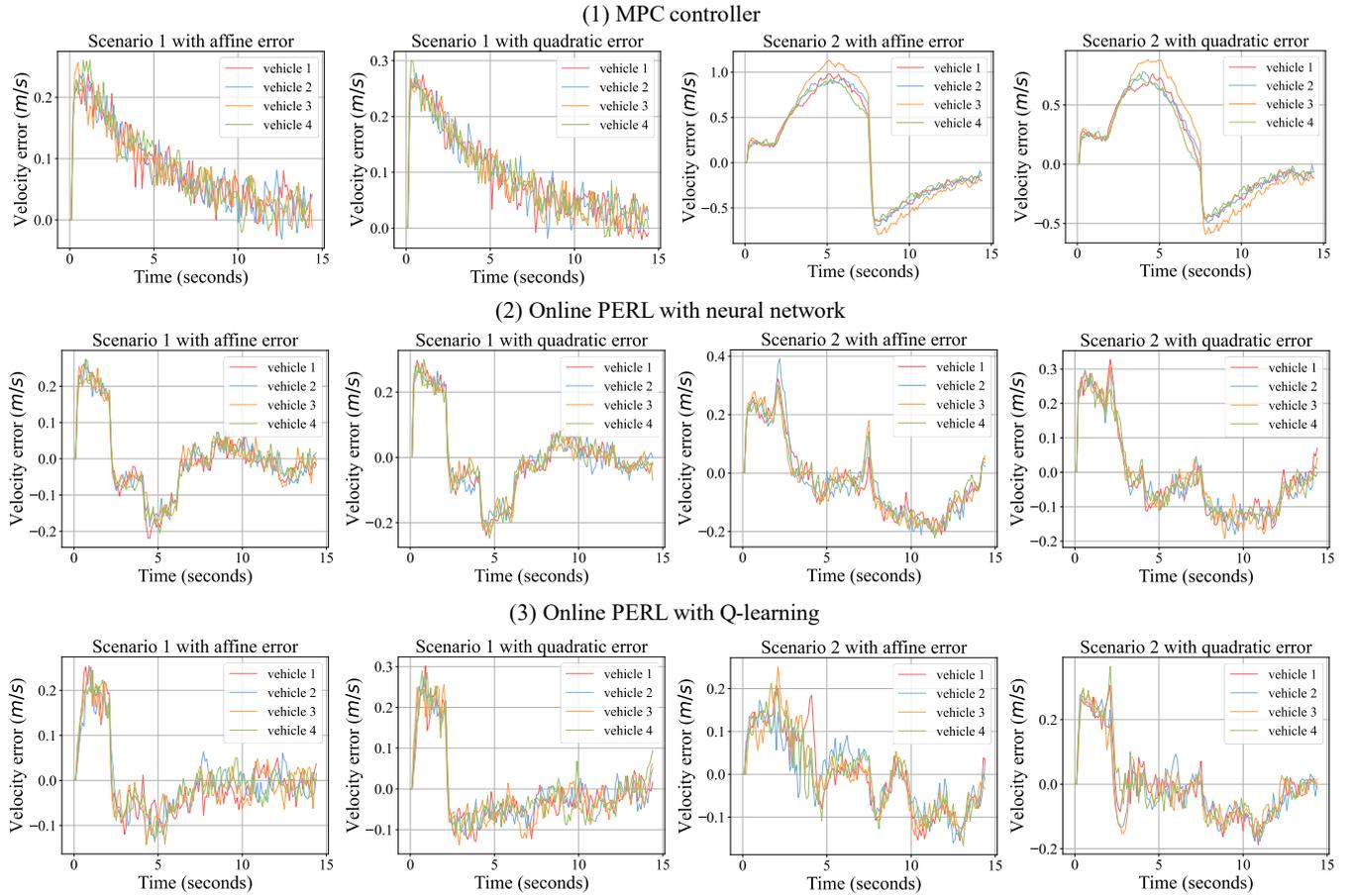

Figure 5. The comparison of experimental results for the three methods.

TABLE 1 Simulation results for the three control models in four tests.

| Test results | Scenario 1 with Affine Error | | | Scenario 1 with Quadratic Error | | | Scenario 2 with Affine Error | | | Scenario 2 with Quadratic Error | | |
|---|---|---|---|---|---|---|---|---|---|---|---|---|
| | M | M+N | M+Q | M | M+N | M+Q | M | M+N | M+Q | M | M+N | M+Q |
| $CAE_p$ | 475.5 | 68.7 | **62.2** | 542.5 | 86.9 | **79.2** | 1388.2 | 199.9 | **161.0** | 1137.0 | 181.5 | **157.0** |
| $GapCAE_p$ | 86.92 | 9.44 | **0.00** | 85.40 | 8.88 | **0.00** | 88.41 | 19.46 | **0.00** | 86.19 | 13.51 | **0.00** |
| $CAE_v$ | 47.9 | 35.0 | **32.9** | 53.9 | 47.9 | **38.6** | 288.0 | 63.3 | **41.7** | 201.1 | 57.4 | **48.4** |
| $GapCAE_v$ | 31.32 | 6.02 | **0.00** | 28.33 | 19.38 | **0.00** | 85.53 | 34.19 | **0.00** | 75.94 | 15.71 | **0.00** |
| $MAE_p$ | 1.129 | 0.421 | **0.370** | 1.356 | 0.493 | **0.417** | 5.220 | 0.616 | **0.493** | 3.978 | 0.618 | **0.552** |
| $GapMAE_p$ | 68.83 | 11.92 | **0.00** | 69.26 | 15.44 | **0.00** | 90.55 | 19.96 | **0.00** | 86.11 | 10.63 | **0.00** |
| $MAE_v$ | 0.260 | 0.267 | **0.254** | 0.301 | 0.300 | **0.302** | 1.132 | 0.391 | **0.251** | 0.884 | 0.327 | **0.312** |
| $GapMAE_v$ | 2.32 | 4.88 | **0.00** | -0.41 | -0.58 | **0.00** | 77.80 | 35.73 | **0.00** | 64.68 | 4.68 | **0.00** |

Notations: $CAE$ and $MAE$ represent the cumulative absolute error and the maximum absolute error. $Gap$ represents the error gap compared with the errors obtained by MPC combined with Q-learning. M, M+N, and M+Q to represent using MPC alone, MPC combined with neural network, and MPC with Q-learning.

## IV. Conclusion

This paper proposes an integrated online PERL framework that incorporates the physical model and residual learning to enhance centralized control of CAVs platoon. To mitigate the disturbance caused during the transition of the control output, the PERL controller applies Q-learning as a residual learning module to adjust the control output of the physical model, i.e., the MPC controller. This integrated model, combining online residual learning with a physical structure, exhibits high precision in predicting control residuals and demonstrates exceptional adaptability. In the MPC controller, the vehicle dynamics' physical model with the inertial delay is incorporated. By setting velocity as the control output, multi-objective optimization under multiple constraints is achieved. For online residual learning, Q-learning is applied to learn the disturbance caused by the complex environment and vehicles' dynamics. The experiments demonstrate that the trajectories by the online Q-learning PERL controller exhibit significantly reduced errors, with cumulative absolute position and velocity errors averaging 86.73% and 55.28% lower than those of the MPC controller, and 12.82% and 18.83% lower compared to the PERL controller utilizing a neural network. The variation in velocity errors highlights the faster convergence speed of the Q-learning in the online learning process.

In the future, we will further explore the suitable physical control model and the residual learning method for the CAV control. Besides, considering that the manual affine and quadric errors cannot demonstrate real-world disturbance, we plan to apply our PERL controller in the real vehicle platform to evaluate the effectiveness and robustness of the controller.